# Environmental Stability of Bismuthene: Oxidation Mechanism and Structural Stability of 2D Pnictogens

Andrey A. Kistanov,[a,*] Salavat Kh. Khadiullin,[b] Kun Zhou,[c, d] Sergey V. Dmitriev[a] and Elena A. Korznikova[a]

Recently a new group of two-dimensional (2D) materials, originating from the group V elements (pnictogens), has gained global attention owing to their outstanding properties. Due to a high surface-volume ratio and extraordinary chemical activity, 2D pnictogens such as phosphorene and antimonene are highly sensitive to exposure to the environment. Hence, upon the exposure to oxygen and water molecules, they may easily oxidize, which leads to the degradation of their structure. In this work, we perform a first-principles investigation on the effects of the environmental oxygen and water molecules on the structural stability of newly emerged group V 2D material bismuthene. It is proposed that the oxidation process of bismuthene and other 2D pnictogens at ambient conditions can involve three general steps: adsorption of oxygen molecules; dissociation of oxygen molecules; interaction of water molecules with the oxygen species anchored at the surface to form acids. Importantly, recent experiments reported on high stability of bismuthene even at high temperatures. Here we show that the underlying reason for such structural stability of bismuthene may have similar roots as the stability of antimonene which originates from an acceptor role of water molecules on that material while in the case of materials with the lower stability, like phosphorene and InSe, water molecules act as donors. The present work uncovers the oxidation mechanisms and suggests the ways for maintaining the stability of bismuthene and its 2D pnictogens counterparts, which may be important for their fabrication, storage, and applications.

## Introduction

After decades of research on graphene and its derivatives such as graphene oxides, graphane, and graphune, the era of new two-dimensional (2D) materials such as transition-metal dichalcogenides[1-3] and MXenes[4,5] has begun. Currently, an avalanche of scientific investigations is directed to the outstanding group of 2D materials - 2D allotropes of pnictogens.[6,7] Recent experimental[8-13] and theoretical[14-21] studies predicted the structures of these materials and proposed a variety of methods for the manufacturing them.

At the first member successfully isolated in the pnictogens' family, phosphorene[22-24] has been widely studied and keeps attracting interest due to its remarkable opto-electronic and mechanical properties[25-35] such as widely tuneable direct gap, strong quantum confinement effect, ultrahigh carrier mobility, and anisotropic Young's. However, a significant obstacle for its application in nanoelectronic devices - its environmental instability: phosphorene degrades upon environmental exposure within hours.[36-40] Following the keen interest in the study of structural stability and properties of phosphorene, current scientific studied are directed on its close counterpart InSe[41,42] and another two group V monolayer elements, arsenene[43,44] and antimonene.[45-47] These pnictogens surpass phosphorene in terms of higher stability.[15,48]

Despite the recent progress in synthesis of highly stable 2D materials,[8,49] because of their large surface area and weak electronic screening, the structure integrity and properties of most 2D materials are still significantly affected by external adsorbates including environmental molecules and dopants.[37,50-57]

For example, it has been shown that the contact of oxygen with the surface of phosphorene negatively affects its stability.[37,50-52] Moreover, the presence of various structural defects significantly enhances the interaction of phosphorene with oxygen, which further promotes its structure degradation.[38,53,54]

Exfoliated monolayer of InSe is susceptible to rapid degradation in ambient conditions, which can be attributed to reactions between the InSe surface and chemical species (water and oxygen) in ambient.[55-57] For instance, exfoliation of InSe in an oxygenated aqueous environment leads to its greater oxidation compared to the case of its exfoliation in a deoxygenated aqueous environment. Thus, oxygen significantly implicates the degradation process of monolayer InSe.[58]

The production and application of arsenene is also challenging due to the formation of arsenic oxides upon its interaction with the environment, which may lead to the lack of properties of arsenene and its structure degradation.[59-61] Nevertheless, the fully oxidized arsenene surface has been predicted to be stable against oxidization and degradation.[62]

Unlike its predecessors, antimonene possesses high chemical stability. High-quality monolayer antimonene grown on a $PdTe_2$ substrate has shown an excellent stability in air exposure.[63] This may be due to highly localized electrons at the Sb–Sb pair, resulting in a strong chemical interaction between each atomic pair.[10]

The latest element belonging to the 2D pnictogens family, bismuthene,[59,64] possess a highly stable hexagonal structure. Depending on the size of the bismuthene sheet, it can vary between a narrow bandgap semiconductor to a metal.[65,66] In addition, differently from other 2D insulators operating at temperatures well below freezing, bismuthene is a room-temperature topological insulator.[65] Therefore, for application of an extraordinary bismuthene in nanoelectronics, including spintronics, computing and data devices, and saturable absorbers, a comprehensive study of bismuthene stability under the ambient conditions as well as the understanding of mechanisms of its interaction with environmental molecules are highly desired. However, the effects of the environmental molecules on the stability of bismuthene have not been explored yet. Moreover, there are not many scientific reports on studying general trends of 2D pnictogens' structure degradation and finding ways for their protection.

In this work, we study the effects of oxygen and water environmental molecules on the electronic properties and structural stability of bismuthene using first-principles calculations. The found strong acceptor role and the low splitting energy barrier (~0.60 eV) of oxygen on bismuthene suggests that at elevated temperatures the formation of bismuthene oxides on its surface is highly possible. Nevertheless, similar to that of antimonene, the stability of bismuthene under the exposure of ambient is maintained due to a donor role of water to the oxidized bismuthene surface. The results obtained here allow the comparison of oxidation and degradation mechanisms taking place in bismuthene and other 2D pnictogens. Accordingly, the strategies for protecting the structural integrity of 2D pnictogens are proposed.

## Computational details

The calculations are conducted in the framework of the density functional theory (DFT) implemented via the Vienna *ab initio* simulation package (VASP).[67] All the considered structures are fully relaxed until the atomic forces and total energy are smaller than 0.01 eV/Å and $10^{-6}$ eV, respectively. To correct the value of the band gap size, which is usually underestimated for the semiconductors by the normal Perdew–Burke–Ernzerhof (PBE) under the generalized gradient approximation (GGA) calculations, the hybrid functional (HSE06)[68] is used for the band structure calculation of pristine bismuthene. The HSE06 calculation predicts a direct band gap of 0.98 eV, while the PBE GGA calculations shows a direct band gap of 0.56 eV. Except the difference in the band gap size, both functionals give qualitatively similar band structures of bismuthene (see Figure 1). Furthermore, the HSE06 and PBE GGA methods predict similar lattice parameters and atomic structures. The relaxed lattice constants of monolayer bismuthene are $a = b = 4.38$ Å (GGA and HSE value), which is in good consistency with previous works.[12,49,69] To avoid a high computational cost of hybrid functional calculations, the PBE GGA calculations are performed for all the considered systems. The effects of molecular adsorbates are considered based on the computational model that includes a single molecule on the bismuthene sheet consisting of a 5×5×1 supercell. A vacuum space of 20 Å is introduced to exclude the interaction between the replicate unit cells. A kinetic energy cut-off of 400 eV is set. The van der Waals-corrected functional Becke88 optimization (optB88)[70] is adopted for consideration of noncovalent chemical interactions between molecules and bismuthene surface.

The adsorption energy ($E_a$) of a molecule on bismuthene is given as

$$E_a = E_{Bi+mol} - E_{Bi} - E_{mol}, \quad (1)$$

where $E_{Bi+mol}$, $E_{Bi}$, and $E_{mol}$ are the energies of the molecule-adsorbed bismuthene, the isolated bismuthene, and the molecule, respectively. The charge transfer between the molecule and the bismuthene surface is given by the differential charge density (DCD) $\Delta\rho(r)$ defined as

$$\Delta\rho(r) = \rho_{Bi+mol}(r) - \rho_{Bi}(r) - \rho_{mol}(r), \quad (2)$$

where $\rho_{Bi+mol}(r)$, $\rho_{Bi}(r)$, and $\rho_{mol}(r)$ are the charge densities of the molecule-adsorbed bismuthene, the isolated bismuthene, and the molecule, respectively. The exact amount of the charge transfer between the molecule and the bismuthene surface is calculated by integrating $\Delta\rho(r)$ over the basal plane at the $z$ point for deriving the plane-averaged DCD $\Delta\rho(z)$ along the normal direction $z$ of the sheet. The amount of transferred charge at the $z$ point is defined as[71]

$$\Delta Q(z) = \int_{-\infty}^{z} \Delta\rho(z')dz'. \quad (3)$$

The nudged elastic band (NEB) method is applied for calculation of the reaction barriers for the oxygen molecule dissociation on the bismuthene surface. The transition time of the $O_2$ molecule from the physisorbed state to the chemisorbed state is estimated as

$$t \approx 1 / f \cdot e^{-E_b/T \cdot k_b} \quad (4)$$

where $E_b$ is the barrier, $k_b$ is the Boltzmann constant, $T$ is a temperature, and $f$ is the attempt frequency. In our calculations, one atmospheric pressure, the room temperature of 300 K, and $f$ of around $10^8$ molecules per second are taken.

*Ab initio* molecular dynamics (AIMD) simulations are performed at room temperature using the Nose-Hoover method with a time step of 1.0 fs.

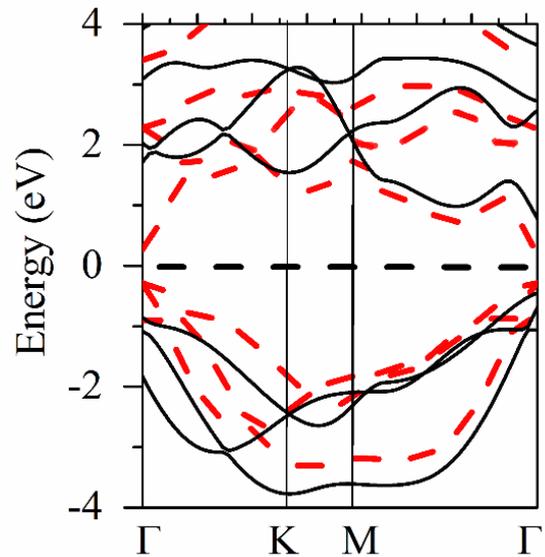

**Figure 1.** The band structure of monolayer bismuthene. The black and red dashed lines show the band structure calculated by HSE06 and PBE GGA methods, respectively. The black horizontal line indicates the Fermi level.

## Results and discussion

**The adsorption of water and oxygen environmental molecules on bismuthene.** It is well known that oxidization taking place upon the interaction of oxygen ($O_2$) and water ($H_2O$) molecules with 2D materials at ambient conditions plays a paramount role in their structure destruction and the decrease of their performances.[55-58] Hence, for bismuthene, it is also important to systematically investigate the effects of the environmental $O_2$ and $H_2O$ molecules on its structural stability, electronic properties and charge transfer ability.

Firstly, the adsorption of the $O_2$ (Figure 2) and $H_2O$ (Figure 3) molecules on the bismuthene surface is considered. Table 1 presents the results about the adsorption energy $E_a$ and the charge transfer $\Delta q$ between the molecules and the bismuthene surface for the lowest-energy configurations. The most energetically favourable configurations together with the DCD isosurface plot for the $O_2$ and $H_2O$ molecules adsorbed on the bismuthene surface are shown in Figures 2a and 3a, respectively. The $O_2$ molecule is located perpendicular to the bismuthene basal plane at the centre of the hexagon with $d = 2.25$ Å.

The DCD plot (Figure 2a) together with the charge transfer analysis (Figure 2b) predicts that the $O_2$ molecule acts as an acceptor (accumulates electrons) to bismuthene with the total $\Delta q = $ ~0.105 $e$ per molecule. The bond length of the $O_2$ molecule significantly increases from 1.22 Å (the free molecule) to 1.27 Å (adsorbed state) upon its adsorption on the bismuthene surface, which may indicate high oxidation ability of bismuthene at ambient conditions.

The LDOS (Figure 2c) and band structure (Figure 2d) plots for the $O_2$ adsorption reflect additional states within the band gap of bismuthene. Despite a slight broadening of the half-filled $2\pi$ HOMO state (Figure 2c), the antibonding $2\pi^*$ LUMO state ($2\pi$, down) is almost unaffected and located within the band gap of bismuthene just above (~0.08 eV) the Fermi level (Figure 2c and d). In addition, the band structure analysis also shows the absence of changes in the band gap size of bismuthene upon the $O_2$ molecule adsorption ($O_2$-induced states are not taken into account). The existence of the $O_2$-induced states within the original band gap of bismuthene together with a strong adsorption ability of the $O_2$ molecule to bismuthene suggests significant changes in the optical and electronic properties of bismuthene upon its oxidation.

For the $H_2O$ molecule, the DCD plot (Figure 3a) together with the charge transfer analysis (Figure 3b) suggests that the molecule donates ~0.041 $e$ to the bismuthene surface. One of the O-H bonds of the $H_2O$ molecule adopts a flat alignment while another one is perpendicular to the bismuthene basal plane with $d = 2.66$ Å.

The LDOS plot for the $H_2O$ molecule (Figure 3c) shows that the $1b_1$, $3a_1$, and $1b_2$ HOMO levels are non-resonant and coincide with the valence states of bismuthene. The $1b_2$ state has the largest orbital mixing with the Bi atom, which allows more efficient charge transfer to the bismuthene surface. The found strong state coupling indicates that the electronic properties of bismuthene may alter in the moisture environment. The band structure analysis (Figure 3d) predicts that the band gap size of the $H_2O$-adsorbed bismuthene is the same as that of perfect bismuthene. In addition, there are no localized states originating from the $H_2O$ molecule within the band gap of bismuthene.

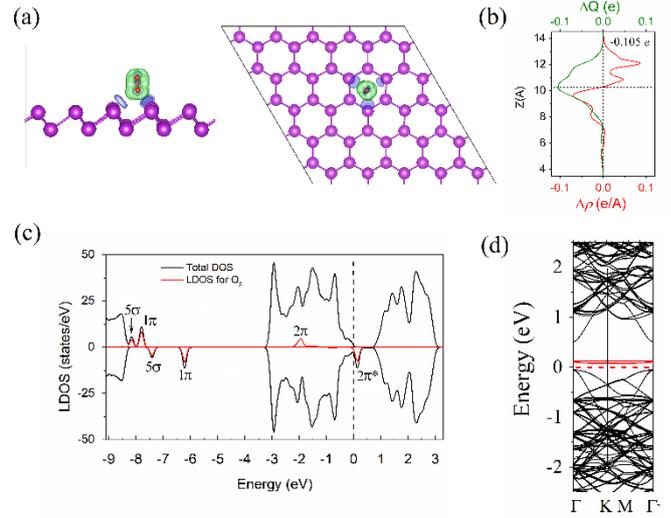

**Figure 2.** (a) The side and top views of the lowest-energy configuration combined with the DCD isosurface plots ($0.8 \times 10^{-3}$ Å$^{-3}$) for the $O_2$-adsorbed bismuthene. The green (blue) colour represents an accumulation (depletion) of electrons. (b) The line profiles of the plane-averaged $\Delta\rho(z)$ (red line) and the transferred amount of charge $\Delta Q(z)$ (green line). (c) The total DOS (black line) and LDOS (red line) of the $O_2$-adsorbed bismuthene, with the black dashed line showing the Fermi level. (d) The band structure of the $O_2$-adsorbed bismuthene, with the red dashed line showing the Fermi level.

The $E_a$ between the $O_2$ molecule and bismuthene is found to be –0.61 eV, which is more than twice that between the $O_2$ molecule and phosphorene ($E_a = -0.27$ eV),[38,71-73] slightly lower than that between the $O_2$ molecule and arsenene ($E_a = -0.54$)[74,75] and equal to that between the $O_2$ molecule and antimonene ($E_a = -0.61$ eV).[76-78] The latter can highly likely be related to the equal electronegativity values of bismuthene and antimonene. In general, one can mention that the whole trend of the charge transfer amount upon oxygen adsorbtion quite accurately repeats the change in the electronegativity value among the studied elements. Furthermore, bismuthene possesses about five times smaller $E_a$ than InSe ($E_a$ = –0.12 eV).[57,79] Considerable differences between interaction characteristics of 2D pnictogens and InSe can be related to the fact that crystals of the indium selenide group have an ionic bonding component[80] that determines the difference in their behaviour from purely 2D covalent crystals.

**Table 1.** Adsorption energy ($E_a$) and the amount of charge transfer ($\Delta q$) between the molecules and the bismuthene surface. Note that a negative (positive) $\Delta q$ indicates a gain (loss) of electrons from each molecule to bismuthene.

| Molecule | InSe[57,79] | | Phosphorene[71] | | Arsenene[81] | | Antimonene[78] | | Bismuthene | |
|---|---|---|---|---|---|---|---|---|---|---|
| | $E_a$ (eV) | $\Delta q$ ($e$) | $E_a$ (eV) | $\Delta q$ ($e$) | $E_a$ (eV) | $\Delta q$ ($e$) | $E_a$ (eV) | $\Delta q$ ($e$) | $E_a$ (eV) | $\Delta q$ ($e$) |
| $O_2$ | -0.12 | -0.001 | -0.27 | -0.064 | -0.54 | -0.076 | -0.61 | -0.116 | -0.61 | -0.105 |
| $H_2O$ | -0.17 | -0.01 | -0.14 | 0.035 | -0.19 | -0.016 | -0.20 | -0.021 | -0.14 | 0.041 |

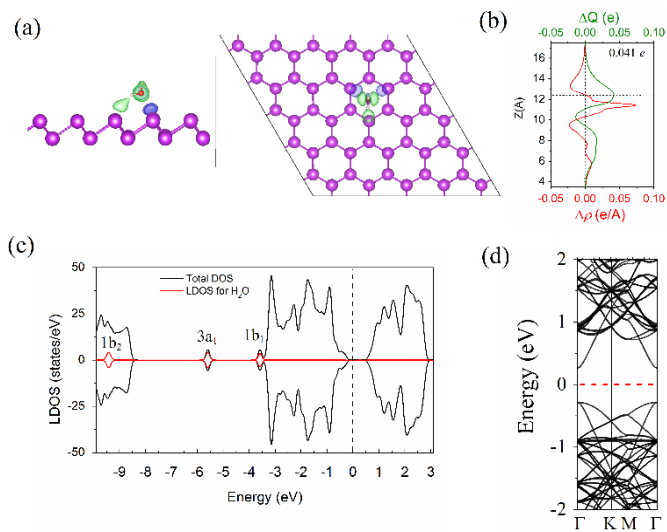

Figure 3. (a) The side and top views of the lowest-energy configuration combined with the DCD isosurface plots (0.8 × 10$^{-3}$ Å$^{-3}$) for the H$_2$O-adsorbed bismuthene. The green (blue) colour represents an accumulation (depletion) of electrons. (b) The line profiles of the plane-averaged $\Delta\rho(z)$ (red line) and the transferred amount of charge $\Delta Q(z)$ (green line). (c) The total DOS (black line) and LDOS (red line) of the H$_2$O-adsorbed bismuthene, with the black dashed line showing the Fermi level. (d) The band structure of the H$_2$O-adsorbed bismuthene, with the red dashed line showing the Fermi level.

The relatively low $E_a$ = –0.14 eV of the H$_2$O molecule on the bismuthene surface suggests a weak interaction between the molecule and the surface. Importantly, a weak interaction with the H$_2$O molecule is found to be common for all group V 2D materials, such as phosphorene ($E_a$ = –0.14),[71] arsenene ($E_a$ = –0.19),[81] antimonene ($E_a$ = –0.20)[78] and their comrade InSe ($E_a$ = –0.17).[57,79]

Similarly, the charge transfer between the O$_2$ molecule and bismuthene (acceptor with $\Delta q$ ~0.105 $e$) is ~30% higher than that between the O$_2$ molecule and phosphorene (acceptor with $\Delta q$ ~0.064 $e$)[71] and that between the O$_2$ molecule and arsenene (acceptor with $\Delta q$ ~0.076 $e$),[81] ~100% higher than that between the O$_2$ molecule and InSe (neutral with $\Delta q$ ~0.001 $e$),[57,79] and slightly lower than that between the O$_2$ molecule and antimonene (acceptor with $\Delta q$ ~0.116 $e$).[78] These results are reasonable since down the group V the electronegativity decreases, which leads to a greater charge transfer to O$_2$ from 2D pnictogens. As shown above, the H$_2$O molecule acts as a donor to bismuthene, which is similar to the role of H$_2$O in the case of its adsorbtion on phosphorene, while for arsenene, InSe, and antimonene the H$_2$O molecule is an acceptor.[78,79,81] Based on the calculated adsorbtion energy and the charge transfer analyses, the performance of 2D pnictogens and InSe is highly sensitive to the environmental O$_2$ molecule rather than the H$_2$O molecule.

Table 2. The energy barrier $E_b$ and the donor/acceptor characteristics of the O$_2$ molecule on the group V 2D materials and InSe.[57,71,78-83]

| Material | $E_b$ (eV) | O$_2$ molecule | H$_2$O molecule |
|---|---|---|---|
| InSe | 1.21-1.24 | weak acceptor | neutral |
| Phosphorene | 0.80 | acceptor | donor |
| Arsenene | 0.67 | acceptor | weak **acceptor** |
| *Antimonene* | 0.24 | strong acceptor | **acceptor** |
| Bismuthene | 0.60 | strong acceptor | donor |
| Pre-oxidized bismuthene | - | - | **acceptor** |

**Oxidation kinetics and mechanisms of the structural degradation of bismuthene.** To understand the underlying oxidation mechanisms, the thermodynamics analysis alone is not enough. Therefore, the kinetic analysis on O$_2$ splitting on the bismuthene surface in the form of terminated –O groups is performed. Figure 4 presents the result from the climbing nudged elastic band (NEB) calculation. The energy barrier $E_b$ for the O$_2$ molecule dissociation on bismuthene is found to be ~0.60 eV, which is much smaller than that for O$_2$ dissociation on InSe[57,82] and on phosphorene,[38,52,54,83] slightly lower than that on arsenene,[35] but larger than that on antimonene[78] (see Table 2).

The calculated low barrier for O$_2$ dissociation on bismuthene suggests its fast oxidation during synthesis and applications at elevated temperatures. According to the rate theory (Eq. 4), the chemisorbtion of the O$_2$ molecule on the bismuthene surface lasts just about 2 min.

Importantly, Figure 4c shows that at the final stage of the O$_2$ dissociation, one of the O atoms protrudes above the bismuthene surface while another one is wedged in the bismuthene surface, which induces a significant structure distortion. Such structural changes upon the O$_2$ molecule dissociation on the bismuthene surface is similar to that of its counterparts: phosphorene, arsenene, antimonene and InSe.[71,78-83]

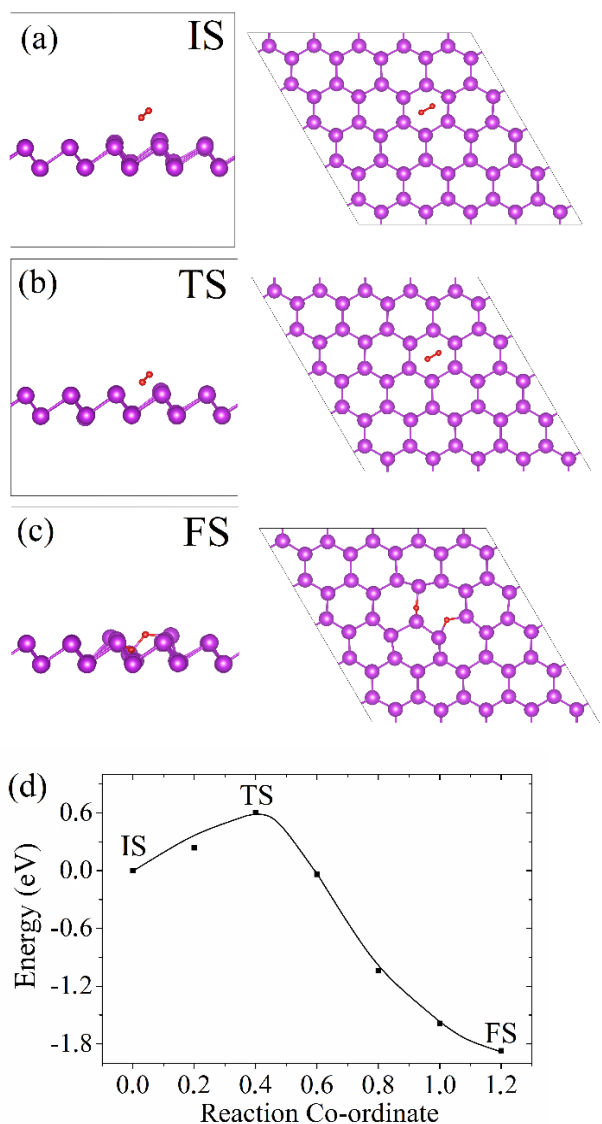

**Figure 4.** The activation barrier for the $O_2$ molecule splitting on bismuthene. (a) The atomic models for the initial state (IS), transition state (TS) and final state (FS). (b) The energy profile obtained by the NEB calculation for the decomposition of the $O_2$ molecule on bismuthene.

**Structure stability/instability of bismuthene and other 2D pnictogens at ambient conditions.** Further, as it has been shown for phosphorene, arsenene and InSe, the oxygen species at the surface tend to react with environmental $H_2O$ molecules, which leads to the degradation of these materials. Taking this theory of degradation of 2D pnictogens as a basis, we investigate the kinetics of the $H_2O$ molecule on two different bismuthene surfaces: i) perfect bismuthene without O groups and ii) partially oxidized bismuthene. In the case of partially oxidized bismuthene, the surface model for the AIMD calculations corresponds to the end states of the NEB calculations shown in Figure 4c.

Our AIMD results suggest that for perfect bismuthene, the $H_2O$ molecule randomly walks above the surface (see movie 1 in SI), which is also supported by the simulated snapshots in Figure 5a. For partially oxidized bismuthene (Figure 5b), the kinetics of the $H_2O$ molecule is significantly different. Initially, the $H_2O$ molecule is located in the vicinity of the terminated O atom in the apical Bi-O group normal to the surface. In the course of events, the $H_2O$ molecule becomes trapped by the O-species on the bismuthene surface and moves around them (see Figure 5b and movie 2 in SI).

In addition, at ~4 ps one of the O-H bonds in the water molecule which is placed closer to the bismuthene surface elongates from 0.96 Å (for the isolated molecule) to 1.026 Å. However, different from InSe and phosphorene, where the formation of the H-O-Se and H-O-P groups on their surface occurs upon a spontaneous dissociation of the water molecule around the terminated O atoms,[38,54] there is no H-O-Bi groups on the pre-oxidized bismuthene surface upon its contact with the water molecule.

Clearly, the roles of the oxygen and/or water molecules and their cooperative effect on the structural stability of bismuthene must be different from those in InSe and phosphorene.

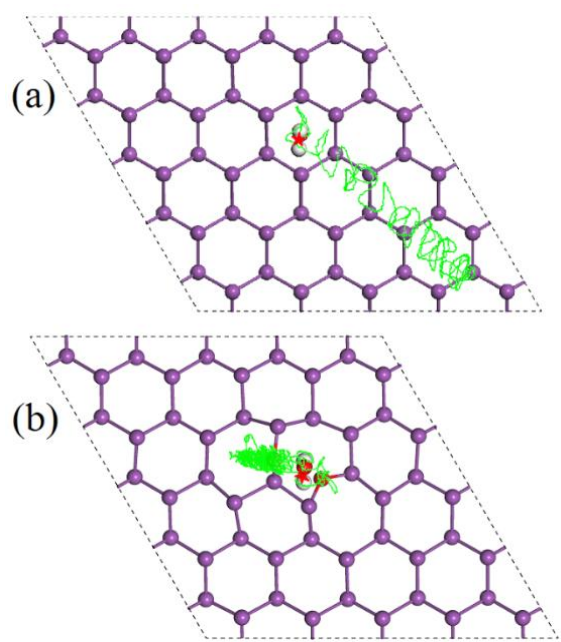

**Figure 5.** The trajectories of the $H_2O$ molecule (O atom) adsorbed on the (a) perfect and (b) pre-oxidized bismuthene surfaces are shown by the green curves. The red stars in the trajectory plots indicate the starting point of the $H_2O$ molecule. The trajectories are shown for the simulation time of ~6 ps at 300 K.

**Degradation mechanisms 2D pnictogens.** The interaction of oxygen molecules with 2D materials plays a critical role in their stability and performance at ambient conditions as oxidization is the most popular form of structural degradation. Here we propose that the oxidation process of 2D pnictogens at ambient conditions can involve three general steps: adsorption of oxygen molecules; dissociation of oxygen molecules; interaction of water molecules with the oxygen species anchored at the surface to form acids.

Indeed, our above-shown results supported by the other experimental and theoretical investigations,[71,78-83] which predicted the low adsorption energy and low barrier for the dissociation of oxygen molecules on 2D pnictogens (and InSe), confirm the validity of our assumptions regarding the first two steps. As it has been shown for InSe and phosphorene, the oxygen species on their surface tend to react with the environmental water molecules, which leads to the formation of acids that entail the fast degradation of these materials.[57,72,82,83] This is well aligned with our predicted third step. On the other hand, the fast oxidation found here of bismuthene is somehow surprising since the recent experimental studies suggested its high stability at high temperatures.[49] Similarly, the latest experimental discovers reported that notwithstanding the high environmental stability of synthesized antimonene, there are always oxygen species above its surface.[9,48]

To understand the underlying reason for such difference in the structural stability of the studied materials we track the behaviour of oxygen and water molecules on their surfaces. According to Table 2 the oxygen molecule plays the same role (acceptor) in InSe and all 2D pnictogens. Therefore, the answer to the question on the stability of 2D pnictogens may be hidden in the role of water molecules. Indeed, Table 2 shows that the water molecule behaves as an acceptor ($[H_2O]^{-\delta}$ with $\delta$ being a small positive real number) for arsenene and antimonene but as a donor ($[H_2O]^{+\delta}$) for phosphorene and bismuthene. However, in the case of pre-oxidized bismuthene (see Figure 6 and Table 2) the water molecule is an acceptor.

It is well known that the formation mechanism of the $H_2CO_3$ acid from the $H_2O$ and $CO_2$ molecules occurs through the diffusion of the $H^{+\delta}$ ion in a partially positively charged $H_2O$ to the negatively charged $-O^{-\gamma}$ group (where $\gamma$ is a small positive real number) in the $CO_2$ molecule. Herein, a similar mechanism based on the electrostatic repulsion between $[H_2O]^{-\delta}$ and $-O^{-\gamma}$ group is proposed to be responsible for the high stability of bismuthene (as well as arsenene and antimonene) and the low stability of phosphorene and InSe. Particularly, in the case of bismuthene (pre-oxidized) and antimonene, the negatively charged $[H_2O]^{-\delta}$ makes the formation and the diffusion of $H^{+\delta}$ proton to the $-O^{-\gamma}$ group unfavourable, while in the case of phosphorene and InSe the electrostatic repulsion between $[H_2O]^{-\delta}$ and $-O^{-\gamma}$ group is absent. Furthermore, with the increase of the atomic number the bond length of 2D pnictogens increases. Accordingly, the transfer of the $H^{+\delta}$ proton to the $-O^{-\gamma}$ group becomes less energetically favourable. Therefore, the stability of 2D pnictogens increases from phosphorene to bismuthene.

Thus, our investigation of the charge transfer behaviours in 2D pnictogens and InSe predicts that the structures for which water molecules act as acceptors tend to be stable under the environmental conditions as there is much less probability for the formation of acids on their surfaces under the co-adsorption of oxygen and water molecules. In addition, our results suggest that the stable surface oxidation layer may be helpful for protecting the underneath 2D pnictogens (arsenene, antimonene and bismuthene) layers. The converse is also true, if the water molecule acts as a donor (or inert), the acids may easily be formed on the surface of 2D pnictogens, which leads to their decomposition as demonstrated for phosphorene and InSe.[16,57,71,79,82]

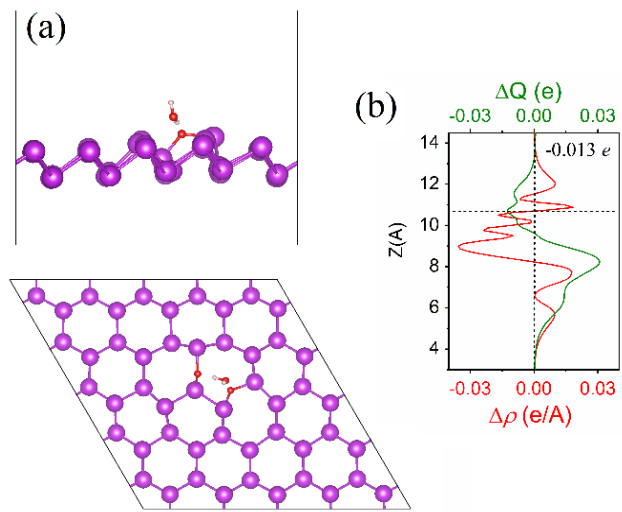

**Figure 6.** (a) The side and top views of the lowest-energy configuration for the $H_2O$ molecule adsorbed on pre-oxidized bismuthene. (b) The line profiles of the plane-averaged $\Delta\rho(z)$ (red line) and the transferred amount of charge $\Delta Q(z)$ (green line).

## Conclusions

DFT-based calculations are conducted to study the energetics and charge transfer of oxygen and water molecules adsorbed on bismuthene. The obtained results are compared with the data for other 2D pnictogens and InSe. The strong acceptor ability of oxygen molecules to bismuthene is found, which is similar to InSe, phosphorene, arsenene and antimonene. The water molecules are predicted to serve as strong donors to bismuthene, similar to phosphorene. However, water molecules play an opposite role (acceptors) to bismuthene in the case of the presence of oxygen species on its surface.

The investigation of the oxygen dissociation kinetics on bismuthene predicts a relatively low barrier of ~0.6 eV for the oxygen molecule decomposition. That is lower than that of phosphorene and InSe, comparable to that of arsenene, and much higher that of antimonene. In addition, such a low barrier for the oxygen molecule splitting on bismuthene suggests its fast oxidation at ambient conditions. Nevertheless, a high stability of bismuthene, similar to arsenene and antimonene, has previously been shown. Here we propose that the underlying reason for that is an acceptor role of water to these materials. In the case where the water molecule acts as an acceptor to 2D pnictogens, the formation of acids on their surface, arising from the interaction between water molecules and oxygen species, is impeded. The opposite is also true when water is a donor to 2D pnictogens; the formation of acids on their surface occurs which leads to their fast degradation.

The present work helps to understand the oxidation process of bismuthene and its 2D pnictogens counterparts and suggests the ways for maintaining their structural integrity under environmental conditions, which is necessary for their successful manufacturing and applications.